\documentclass[fleqn,usenatbib]{mnras}
\usepackage[T1]{fontenc}
\usepackage{ae,aecompl}
\usepackage{graphicx}   % Including figure files
\usepackage{amsmath}    % Advanced maths commands
\usepackage{amssymb}    % Extra maths symbols
\usepackage{float}
\usepackage[normalem]{ulem}

\usepackage[dvipsnames]{xcolor}
\usepackage{orcidlink}

\title[Magnetic field and plasma  
density in AGN jets]
{Magnetic field and plasma number density from radio and millimeter core measurements in AGN jets}
\author[E.E. Nokhrina et al.]{\parbox{\textwidth}{
E.~E.~Nokhrina$^{1,2}$\thanks{E-mail: nokhrinaelena@gmail.com}, A.~P.~Lobanov$^{4}$, A.~Yu.~Istomin$^{1}$,
V.~A.~Frolova$^{2,3}$
}
\vspace{0.4cm}\\
\parbox{\textwidth}{
$^1$Division of Theoretical Physics, P.N. Lebedev Physical Institute, Leninsky prosp.~53, Moscow, 119991, Russia \\
$^2$Relativistic Astrophysics Laboratory, Moscow Institute of Physics and Technology, Institutsky per. 9, Dolgoprudny 141700, Russia \\ 
$^3$The Institute for Nuclear Research of the Russian Academy of Sciences, 60th October Anniversary Prospect 7a, Moscow 117312, Russia \\
$^4$Max-Planck-Institut f\"{u}r Radioastronomie, Auf dem H\"{u}gel 69, Bonn, 53121, Germany
}
}

\begin{document} 

\date{{Accepted 2026 June 30. Received 2026 June 11; in original form 2026 March 13}}

\pagerange{\pageref{firstpage}--\pageref{lastpage}} \pubyear{}

\maketitle

\label{firstpage}

  \begin{abstract}
  Understanding the mechanism for launching relativistic jets in active galactic nuclei relies upon measuring the magnetic field strength and emitting plasma number density, tracing their evolution along the jet, and determining the relation between their rest frame energy densities. This can be achieved using measurements of the size and brightness temperature of the compact region at the jet base (the ``core'') obtained with very long baseline interferometry (VLBI) across frequencies from 2 to 230~GHz.
  We develop a framework for independently estimating the magnetic field $B_*$ and the emitting plasma number density $N_*$ as functions of the jet width $d$, using multifrequency VLBI observations of the core size and brightness temperature. 
   We apply the standard model of self-absorbed synchrotron emission, assuming power-law dependencies of the jet Doppler factor, Lorentz factor, magnetic field strength, and plasma density on the jet width. For an arbitrary jet boundary shape, we derive the dependencies $B_*(d)$ and $N_*(d)$,  
   and explore a possible relation between the rest frame energy densities of the magnetic field and the emitting plasma.
  Analysis of core widths and brightness temperatures measured at multiple frequencies points to the possible presence of a magnetic flux decay and effective plasma acceleration within the observed cores at least in some sources of the sample.
  \end{abstract}


   \maketitle
%
%-------------------------------------------------------------------

\section{Introduction}
\label{s:intro}

Relativistic jets launched in active galactic nuclei (AGNs) were first discovered over a hundred years ago \citep{Curtis1918}, and since then tremendous progress has been made in understanding their physics \citep[see the reviews by][]{Beskin10, BMR-19}. Analytical and numerical magnetohydrodynamic (MHD) modelling provided general understanding of jets as transversally stratified axisymmetric outflows carrying both Poynting and plasma kinetic energy fluxes \citep[e.g.,][]{ChLB-91, AC-92, Eichler-93, Bog-95, VK-03, Vlahakis04, Beskin06, Kom07, McKinney06, Beskin09, Kom09, Lyu09, TMN09, Beskin10, Tchekhovskoy_11, McKinney12}. These works offered valuable insights into the magnetic field structure and bulk acceleration of the jet plasma. In particular, the most generic MHD relations indicate that the toroidal magnetic field component dominates throughout most of the jet \citep[see, e.g.,][and the references within]{Lyu09, Beskin10}. Analysis of the asymptotic behaviour of the MHD equations shows, however, that the poloidal component prevails in the plasma rest frame within the acceleration and collimation zone \citep{Kom07}. Since the toroidal and poloidal components exhibit such distinct properties, determining their relative prevalence becomes crucial for modelling jet emission. 

Jet emission at radio and millimetre wavelengths is well explained by self-absorbed synchrotron  radiation from relativistic plasma in the presence of a magnetic field. However, MHD modelling provides only the total plasma number density in a jet, while only a fraction of the plasma may be relativistic and therefore radiating \citep{Lyutikov2005, Fromm2022, CruzOsorio2022, Frolova23, Hir25}. This makes it challenging to directly connect MHD modelling results and observations. Core shift measurements can provide information on both the magnetic field, $B_*$, and the number density of the emitting plasma, $N_*$, in the plasma rest frame, but such estimates require an additional assumption regarding the relationship between their energy densities \citep{L98, hirotani2005, NBKZ15, N24, N24b}. Consequently, an independent observational determination of $B_*$ and $N_*$ has not yet been achieved. Therefore, the rates at which the  magnetic field strength and plasma density decrease along a jet are typically adopted from theoretical models \citep[see, e.g.,][]{BlandfordKoenigl1979, L98, N24}. 

In this paper we present a method for independently constraining $B_*(d)$, $N_*(d)$ and the relation between their energy densities using VLBI measurements of the core width and a brightness temperature at multiple frequencies. We apply this method under the most general assumptions to the recent results from the Event Horizon Telescope (EHT) Collaboration \citep{Roder25}, noting that it can be equally well applied for analysis of other compilations of multifrequency VLBI measurements of brightness temperature \citep[e.g.,][]{Lee2016, Nair2019}. We also show that complementing this approach with measurements of core shifts can yield a more complete picture of the jet structure within acceleration and collimation zone.

%--------------------------------------------------------------------
\section{Observed and modelled quantities}
\label{s:First}

\subsection{Model assumptions}
\label{ss:model_assump}

It is convenient to express the physical quantities characterising the basic properties of plasma and electromagnetic fields in jets as functions of the local jet width $d$.

In the standard stationary MHD framework, the poloidal ($B_\mathrm{p}$) and toroidal ($B_\varphi$) magnetic field components are defined as 
\[
B_\mathrm{p}=(\nabla\Psi\times\hat{e}_{\varphi})/\pi d, \quad B_{\varphi}=-(4I/cd)\,\hat{e}_{\varphi}\,,
\] 
where $c$ is a speed of light and $\hat{e}_{\varphi}$ is a unit vector in a $\varphi$ (toroidal) direction.
In well collimated flows, both the gradient of a magnetic flux function $\Psi$ and the electric current $I$ vary only weakly with distance $r$ along the jet, so that $B_\mathrm{p}$ and $B_{\varphi}$ primarily depend on the jet width $d$: $B_\mathrm{p}\propto d^{-2}$ and $B_{\varphi}\propto d^{-1}$ \citep{BM00, Beskin06}. The commonly adopted conditions of continuity and conservation of the emitting plasma number density $N$ and the total magnetic flux $\Psi_\mathrm{tot}$ likewise make the jet width $d$ a natural argument for $N$ and $B_\mathrm{p}$. 

MHD modelling further shows that the bulk Lorentz factor, $\Gamma$, of the plasma depends on the jet's axial radius \citep[see, e.g.,][]{Beskin10}. In the magnetically dominated part of the jet, this dependence follows the linear acceleration relation
\begin{equation}
\Gamma=\frac{d}{2R_\mathrm{L}}\,,
\label{eq:gamma_d}
\end{equation}
where the light cylinder radius $R_\mathrm{L} = c/\Omega_0$, defined by the characteristic angular velocity $\Omega_0$, serves as a fundamental length scale.

Using the jet width $d$ as the independent variable (instead of the distance $r$ along the jet) is more convenient for non-conical outflows, for which the $d(r)$ dependence becomes non-linear. 
In the case of radio and millimetre core observations considered in this work, the cores are expected to be located in non-conical regions of the jet. Indeed, in most sources where a transition from a parabolic (or quasi-parabolic) to a conical shape has been observed, the radio cores lie within the quasi-parabolic domain \citep[see, e.g.,][]{Nakamura+18, N24b}.

We therefore introduce a power-law dependence of the jet boundary shape on distance $r$:
\begin{equation}
d\propto r^{k}.
\label{shape}
\end{equation}
Without loss of generality, we write the corresponding expressions for the magnetic field, $B_*$, and the emitting plasma density, $N_*$, in the plasma rest frame as
\begin{equation}
B_*\propto d^{-b}\,,
\label{B}
\end{equation}
\begin{equation}
N_*\propto d^{-s}\,.
\label{n}
\end{equation}
The indices $b$ and $s$ can be related to the conventional notations \citep{L98, Roder25} through $B_*\propto r^{-m}$, $m=k\,b$, and $N_*\propto r^{-n}$, $n=k\,s$.

The physical quantities we aim to determine also depend on the Doppler factor of the emitting plasma 
$\delta = \Gamma^{-1}\,(1-\beta \cos\theta_v)^{-1}$, for a bulk motion at a speed $v=\beta c$ in the direction making an angle $\theta_v$ with the observer's line of sight.
So far, analytical models used to interpret observations of ultracompact jets have typically assumed constant Lorentz factors and, consequently, constant Doppler factors \citep[e.g.,][]{BlandfordKoenigl1979,marscher1990,lobanov2000,hirotani2005,Sullivan09_coreshift,Nair2019}, while fewer studies have explicitly accounted for variations of the Lorentz factor along the jet\citep[e.g.,][]{LZ1999,Lee2016}. However, both observational \citep{Hada18, Ricci22} and theoretical \citep{Kom07, Lyu09, Kov-20} evidence indicate that in the parabolic domain of the jet, where most radio cores are located, the plasma continues to accelerate efficiently following Equation~(\ref{eq:gamma_d}). Bearing this in mind, we introduce a variable Doppler factor for the bulk plasma motion:
\begin{equation}
\delta\propto d^{\,t},
\label{delta}
\end{equation}
or 
$\delta=\delta_0(r/R_\mathrm{L})^{tk}$.
Introducing the maximum Lorentz factor for a given distance $r$ along a jet as $\Gamma=\Gamma_0(r/R_\mathrm{L})^k$, we find for the Doppler factor exponent $t$ the following:
\begin{equation}
t = -\,\frac{%
\ln\;\Bigl[\delta_{0}\,\Bigl(\Gamma_{0} (r/R_\mathrm{L})^{k}
       -\cos\theta_v\,\sqrt{\Gamma_{0}^{2}(r/R_\mathrm{L})^{2k}-1}\Bigr)\Bigr]
}{k\,\ln (r/R_\mathrm{L})}\,.
\label{t_prcise}
\end{equation}
Assuming that the observer's line of sight lies inside an emission cone with the opening angle $\theta_\mathrm{em}=1/\Gamma$, i.e., $\Gamma_{0} (r/R_\mathrm{L})^{k}\ge 1/\theta_v$ for the observed part of the jet, we obtain
$$
t=1-\frac{\ln\;\Bigl(\delta_{0}/\Gamma_{0}\Bigr)}{k\,\ln (r/R_\mathrm{L})}.
$$
For $\delta_0=\Gamma_0$, which holds for $\theta_v=1/\Gamma_0$, the resulting exponent $t=1$. If the emission is dominated by the jet region with $\Gamma\approx 1/\theta_v$, then $\delta\approx \mathrm{const}$ can be assumed for the modelling. Thus, below we regard two limiting cases: $t=0$ corresponds to a constant Doppler factor, whereas $t=1$ represents the limit $\delta\approx \Gamma$, which holds when the viewing angle satisfies the condition $\theta_v\lesssim\Gamma^{-1}$ for all Lorentz factors relevant to the observed cores.

\subsection{Observed quantities}
\label{ss:observables}

The primary observables used in this paper are the multifrequency VLBI measurements reported in \citet{Roder25} for a sample of jet cores represented by their angular sizes, $\theta_\mathrm{obs}$ (reciprocating the jet width $d$) and brightness temperatures, $T_\mathrm{br,obs}$, and described as a function of the observing frequency $\nu$:
\begin{equation}
d\propto\nu^{-p}\,,
\label{d-nu}
\end{equation}
\begin{equation}
T_\mathrm{br,\,obs}\propto\nu^{-q}\,,
\label{T-nu}
\end{equation}
with the cosmological redshift corrections applied to $\nu$ and $T_\mathrm{br,\,obs}$ quantities to convert them to the respective source rest frames. Throughout this paper we adopt the power-law indices $p$ and $q$ obtained by \citet{Roder25}.

We relate $d(\nu)$ and $T_\mathrm{br,obs}(\nu)$ to the distance $r$ along the jet by assuming that the observed location of the core is set by synchrotron self-absorption \citep{BlandfordKoenigl1979}, such that 
\begin{equation}
r\propto \nu^{-1/k_\mathrm{r}}\,,
\label{rcore}
\end{equation}
where the index $k_\mathrm{r}$ characterizes the magnitude of the ``core shift'' with frequency \citep{Koenigl1981,Marcaide1984,L98}. The value $k_\mathrm{r}=1$ is commonly associated with the equipartition condition in the jet \citep{BlandfordKoenigl1979}.
For jets with small local opening angles, $d(r)/r\ll 1$, substituting Equations~(\ref{shape})--(\ref{delta}) into Equation~(4) from \citet{N24}, yields a general expression for the core shift index $k_\mathrm{r}$:
\begin{equation}
k_r=k\frac{s+(1.5-\alpha)(b-t)-1}{2.5-\alpha}\,,
\label{kr}
\end{equation}
where $\alpha$ is the spectral index of the synchrotron emisison, defined as $S_{\nu}\propto\nu^{\alpha}$. This equation is identical to Equation~(13) from \citet{N24}. 
For the equipartition case \citep{BlandfordKoenigl1979}, characterised by $s=2$, $b=1$,  and a conical ($k=1$), non-accelerating ($\delta=\mathrm{const}$, $t=0$) jet, Equation~\ref{kr} gives the expected value $k_r=1$. The general form of $k_\mathrm{r}$ described by Equation~\ref{kr} will be considered throughout this paper.

\subsection{Magnetic field strength and plasma number density}

The brightness temperature measurements discussed above can be used to estimate the magnetic field strength in the rest frame \citep{Nok17A,Roder25} 
\begin{equation}
B_*\propto\nu\delta T_\mathrm{br,\,obs}^{-2}\,,
\label{B-Tbr}
\end{equation}
We eliminate the observing frequency $\nu$ from Equations~(\ref{d-nu})--(\ref{T-nu}), obtaining $T_\mathrm{br,\,obs}\propto d^{q/p}$, and then use Equations~(\ref{B}) and (\ref{delta}) to rewrite the proportionality in Equation~(\ref{B-Tbr}) as
\begin{equation}
d^{-b}\propto d^{\,t-(1+2q)/p}\,,
\end{equation}
which allows us to express the index $b$ in terms of the observed $p$, $q$ and the well-constrained Doppler factor index $t$:
\begin{equation}
b=\frac{1+2q}{p}-t\,.
\label{b}
\end{equation}
Combining the jet boundary and core position relations given by Equations~(\ref{shape}) and (\ref{rcore}),
we obtain
\begin{equation}
\frac{k_r}{k}=\frac{1}{p}\,,
\label{kr-k}
\end{equation}
which can be used together with Equations~(\ref{kr}) to derive an expression for the index $s$ as a function of $p$, $q$, and $t$:
\begin{equation}
s = 1 + \frac{1}{p} + (3-2\alpha)\left(t - \frac{q}{p}\right)\,.
\label{s}
\end{equation}
With $p$ and $q$ constrained by the measured dependencies $T_\mathrm{br,obs}(\nu)$ and $d(\nu)$, Equations~(\ref{b}) and (\ref{s}) provide estimates of the indices $b$ and $s$, which describe how $B_*$ and $N_*$ scale with the jet width $d$, for any assumed value of $t$. 

Equations~(\ref{b}) and (\ref{s}) allow one to estimate the power-law indices $b$ and $s$ describing the dependence of $B_*$ and $N_*$ on the jet width $d$, for any assumed value of $t$. This requires only estimates of the indices $p$ and $q$, which can be obtained from multifrequency measurements of the core width and brightness temperature for either individual sources or source samples.

\section{Application to multifrequency data on jet cores}

\subsection{Specific model scenarios}
\label{sc:scenarios}

We now consider the expected values of indices $p$ and $q$ for different outflow regimes in the jet. The general expressions for $p$ and $q$, obtained from the Equations~(\ref{b}) and (\ref{s}) are
\begin{align}
p &= \frac{2.5-\alpha}{s+(1.5-\alpha)(b-t)-1},
\label{P} \\
q &= \frac{(2.5-\alpha)(b+t)}{2s+(3-2\alpha)(b-t)-2}-\frac{1}{2}.
\label{Q}
\end{align}
In the following, we set $\alpha=-0.5$ noting that the results depend only weakly on the assumed value of $\alpha$.

\begin{table}
      \caption[]{Parameters of physical scenarios}
         \label{t:PQ} 
         $$
         \begin{array}{cccc}
            \hline
            \noalign{\smallskip}
            Scenario & (b,\,s,\,t) &  p & q \\
            \noalign{\smallskip}
            \hline
            \noalign{\smallskip}
            1 & b=2,\,s=3,\,t=1& 0.75 & 0.625   \\
            2 & b=2,\,s=4,\,t=1 & 0.6 & 0.4 \\
            3 & b=2,\,s=3,\,t=0 & 0.5 & 0 \\
            4 & b=2,\,s=4,\,t=0 & 0.43 & -0.1 \\
            5 & b=1,\,s=2,\,t=0 & 1 & 0 \\
            6 & b=3,\,s=1,\,t=1 & 0.75 & 1.0 \\
            \noalign{\smallskip}
            \hline
         \end{array} 
         $$
   \label{t:scen}
   \end{table}

Table~\ref{t:PQ} lists the values of $p$ and $q$ for several representative scenarios corresponding to specific combinations of the model parameters $b$, $s$, and $t$. Scenarios~1 and~2 describe the acceleration region of a jet with a dominant poloidal magnetic field component ($b=2$), a sufficiently small viewing angle such that $\delta\sim 2\Gamma$ ($t=1$), and either particle continuity and conservation along the jet (scenario 1) or equipartition between the emitting plasma and the magnetic field $B_*$ (scenario 2). Scenarios 3 and 4 are analogous, but correspond to a larger viewing angle, so that $\delta\approx\mathrm{const}$. Scenario 5 represents a region of saturated acceleration ($\Gamma\approx\mathrm{const}$, $t=0$) with a dominant toroidal magnetic field ($b=1$) and equipartition ($s=2$) between the emitting plasma and magnetic field energy density in the rest frame of the flow. The last scenario 6 yields indices close to those obtained in Section~\ref{ss:multi_data}. It has not been considered before, as the value $b=3$ requires substantial deviation from a helical magnetic field with conserved magnetic flux, expected from theoretical models \citep{BlandfordKoenigl1979, L98, N24}. Furthermore, the value of index $s=1$ implies a much slower decrease of emitting plasma number density with a jet width, than in the case of the conserved particle flux.

\begin{figure}
    \centering
    \includegraphics[width=1.0\linewidth]{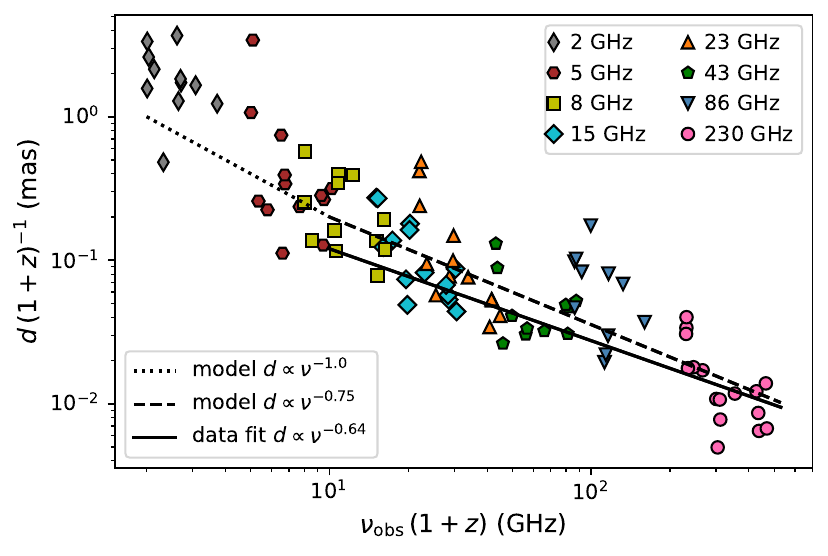}
    \caption{Comparison of model scenarios~1 and~5 (see Table~\ref{t:scen} and Sect.~\ref{sc:scenarios}) for the expected evolution of the core size as a function of the observing frequency $\nu_\mathrm{obs}$. The data points and solid lines are taken from \citet{Roder25}. The dotted and dashed lines show model scenario 5 at $\nu_{\mathrm{obs}}=1$--10~GHz and model scenario~1 at $\nu_{\mathrm{obs}}=10$--500~GHz, respectively.}
    \label{f:p}
\end{figure}

\begin{figure}
    \centering
    \includegraphics[width=1.0\linewidth]{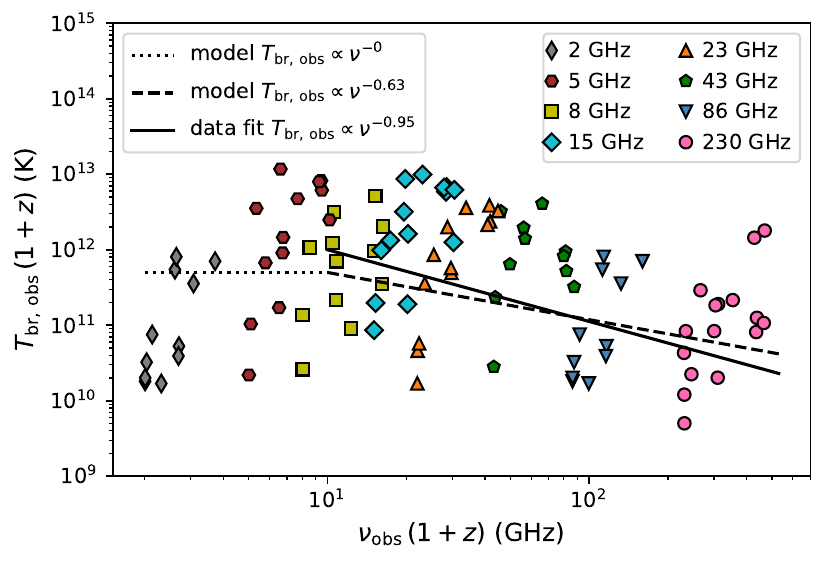}
    \caption{The same as Figure~\ref{f:p}, but for the observed brightness temperature.}
    \label{f:q}
\end{figure}

Figures~\ref{f:p} and \ref{f:q} illustrate the expected behaviour of the observed core size $d$ and brightness temperature $T_\mathrm{b,\,obs}$ for an accelerating jet at small distances from the black hole (high observing frequencies; dashed lines; scenario 1) and for a jet with an approximately constant bulk Lorentz factor at lower frequencies (dotted lines; scenario 5). In this classical picture, the core size is expected to exhibit a break, while the brightness temperature remains approximately constant at lower frequencies, in the plasma-dominated regime. This model behaviour qualitatively reproduces the principal trends reported by \citet{Roder25} and shown by the data points and solid lines in Figures~\ref{f:p} and \ref{f:q}.

\subsection{Multifrequency data}
\label{ss:multi_data}

The relations given by Equations~(\ref{b}) and~(\ref{s}) can be applied for interpreting the results reported by \citet{Roder25} from VLBI measurements of sizes and brightness temperatures of the cores of compact extragalactic jets obtained at multiple frequencies in the range from $2$ to $230$~GHz. As these data results from observations made at different epochs, it is not feasible to use them for separately analysing individual sources in which the emission variability can easily result in more than an order of magnitude scatter of brightness temperature estimates \citep[see, e.g.,][]{Kovalevetal2005,Lee2008,Nair2019}. We therefore resort to using the combined data from \citet{Roder25} in our analysis, noting that although these data are still affected by the variability induced scatter, its negative effect should be alleviated by the overall increase of the number of measurements available for the modelling.

\begin{figure}
    \centering
    \includegraphics[width=1.0\linewidth]{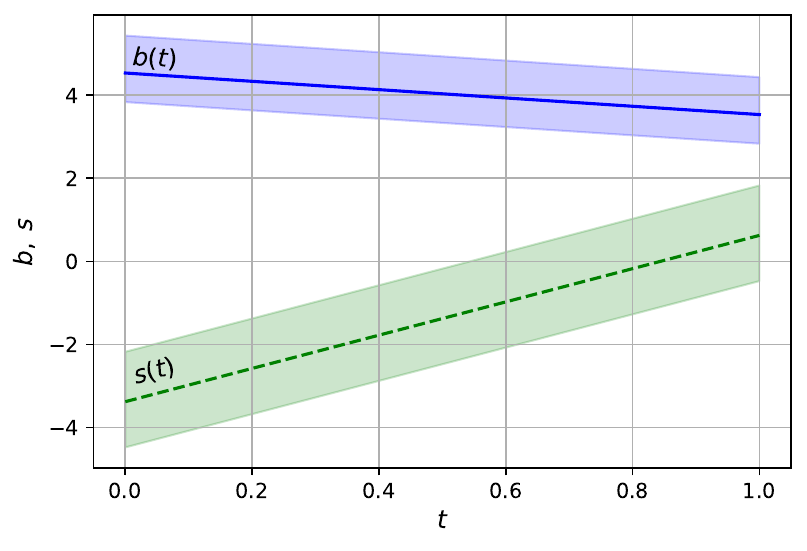}
    \caption{
Dependence of the indices $b$ (blue solid line) and $s$ (green dashed line), describing the scaling of $B_*$ and $N_*$ with jet width $d$, on the Doppler-factor evolution index $t$. The relations are calculated from Equations~(\ref{b}) and~(\ref{s}) using the values of $p$ and $q$ reported by \citet{Roder25} and assuming $\alpha=-0.5$. The shaded regions reflect the reported uncertainties in $p$ and $q$.}
    \label{f:bs_t}
\end{figure}

For the analysis, we adopt $p=0.64\pm 0.05$ and $q=0.95\pm 0.13$ from \citet{Roder25}, and assume a spectral index $\alpha=-0.5$. Under these conditions, Equations~(\ref{b}) and (\ref{s}) yield 
\[
b=\left(4.5^{+0.9}_{-0.7}\right)-t\,,\quad  s=4t-\left(3.4^{+1.2}_{-1.1}\right)\,.
\] 
These dependences are illustrated in Fig.~\ref{f:bs_t}. In the following, we consider two specific scenarios for the evolution of the Doppler factor, $\delta$, corresponding to $t=0$ and $t=1$.

For all values of $t\in [0,1]$, the index $b$, which characterizes the decline of the magnetic field $B_*$ with the jet width, significantly exceeds $2$, the maximum value expected for a predominantly poloidal magnetic field under the assumption of conservation of the total magnetic flux $\Psi_\mathrm{tot}$. Analytical modelling shows that values of $b$ satisfying the condition $b=2+\varepsilon$ may arise along a given magnetic surface owing to a particular, core-like poloidal field structure \citep{N25}, 
leading to a slightly steeper than $\propto d^{-2}$ decrease of the poloidal component of the magnetic field, with $\varepsilon\lesssim 0.3$.

Larger deviations from $b=2$ can be explained only by a decay of the total magnetic flux along the jet, with $\Psi_\mathrm{tot}(d)\propto d^{-\varepsilon}$. For a non-conical flow, this translates into a longitudinal scaling $\Psi_\mathrm{tot}(r)\propto r^{-k\varepsilon}$. Two possibilities may account for such behaviour.

First, the total magnetic flux may dissipate physically along the jet, for example as a result of developing instabilities and subsequent magnetic reconnection. Alternatively, the apparent flux decay may arise if the observationally brightest region of the outflow shifts to domains with progressively smaller magnitude of $\Psi$ with increasing distance along the jet. In this case, one would infer a magnetic field decline steeper than $b=2$. Both mechanisms can be viable over a limited range of distances along the jet. 

The corresponding index $s$, which characterizes the dependence of the emitting plasma number density $N_*$ on the jet width, attains its maximum value, $s=0.6\pm 0.9$, for $t=1$. 
More generally, $s$ is expected to lie in the range of 2--4. In a non-accelerating jet with a constant bulk motion Lorentz factor \citep[e.g.,][]{BlandfordKoenigl1979}, the continuity and energy density conservation conditions for the emitting plasma imply $s_0=2$.

In an accelerating jet, the continuity relation includes variations of the Lorentz factor along the jet, yielding $s_0=3$.
If, in the rest frame of the emitting plasma, the jet is dominated by a poloidal magnetic field component and energy density equipartition holds, one obtains $s_0=4$.

Consider the deviation of the index $s$, assessed from the observational data, from the value $s_0$ described above: $s=s_0-\epsilon$. Negative values of $\epsilon$ yield $s>s_0$ and correspond to a more rapid decrease of emitting plasma number density than expected from the conservation of $N_*$ along the jet ($s_0=2$ for a flow with a constant Lorentz factor or $s_0=3$ for a linearly accelerating flow, \citet{N24}) or from equipartition with the magnetic field ($s_0=2$ for a flow with a constant Lorentz factor or $s_0=4$ for a linearly accelerating flow). Thus, a negative $\epsilon$ implies a decay of emitting plasma density, possibly due to cooling..

A positive value of $\epsilon$ indicates an effective supply of emitting plasma along the jet. Such a supply may arise if a fraction of the cold plasma is continuously accelerated to ultra-relativistic energies.

Therefore, observationally inferred values of $s$ of the form $s=3-\epsilon$ support a continuous injection of emitting plasma along the jet, with a rate of 
\[
\dot{N}\propto\Gamma N_* d^2\propto d^{\,\epsilon}\propto r^{\,k\epsilon}\,.
\]

\begin{figure}
    \centering
    \includegraphics[width=1.0\linewidth]{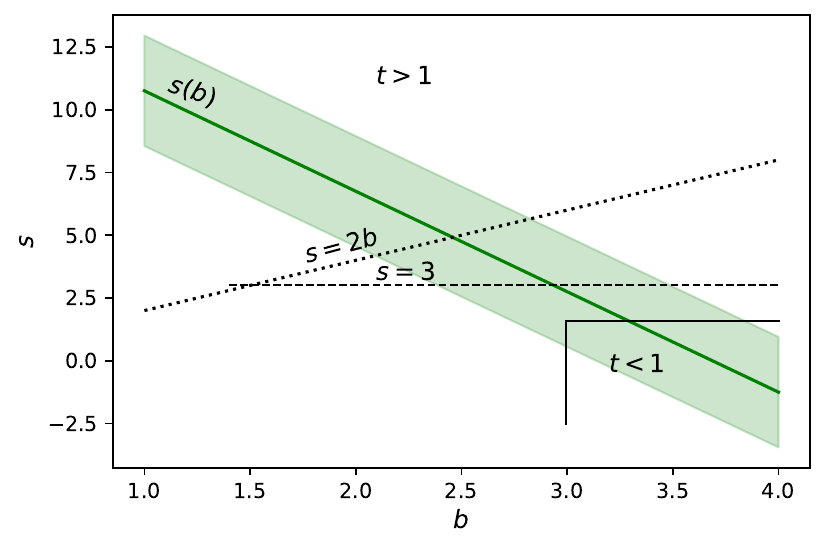}
    \caption{Relation between the power indices $b$ and $s$ (green line) and its uncertainty (green shaded region), computed for the measured values of of $p$ and $q$. The dashed black line denotes the condition of the emitting plasma number density conservation and continuity along a linearly accelerating jet with $\Gamma\propto d$ \citep{N24}. The dotted black line corresponds to equipartition between the emitting plasma and the magnetic field energy densities. The solid black line separates the regions with $t<1$ (consistent with the MHD acceleration rate of the plasma bulk) and $t>1$ (inconsistent with the maximum MHD acceleration rate).}
    \label{f:s_b}
\end{figure}

\noindent
The inferred deviations from the expected ranges $b\in[1,2]$ and $s\in[2,4]$ may also result from averaging over the estimates obtained for different sources. For example, the alternative values $p'=0.72$ and $q'=0.46$ based on a different averaging procedure employed in \citet{Roder25}, yield $b=1.7$ and $s=3.8$, closer to both equipartition and emitting plasma continuity and conservation. 
 
Eliminating the Doppler factor index $t$ from Equations~(\ref{b}) and (\ref{s}) gives a direct relation between $s$ and $b$:
\begin{equation}
s=(3-2\alpha)\left(\frac{1+q}{p}-b\right)+1+\frac{1}{p}.
\label{Eq_s_b}
\end{equation}
Let us consider a broad range of possible values of $b$, from $1$ (purely toroidal component) to $2$ (purely poloidal field), and up to $3.5$, which would imply magnetic flux dissipation along the jet.
The resulting dependence  of $s$ on $b$, computed for the measured $p$ and $q$ and assuming $\alpha=-0.5$ is shown in Fig.~\ref{f:s_b}. 

The solid black line in Fig.~\ref{f:s_b} divides the plane into regions compatible with MHD bulk acceleration rate $\Gamma\propto d$ of the emitting plasma. For sufficiently small viewing angles, the relation $\delta\approx 2\Gamma$ holds, implying a maximum value of $t=1$ (lower right region in Fig.~\ref{f:s_b}). The remaining part of the $b$--$s$ plane corresponds to $t>1$, which cannot be achieved in any viewing geometry without invoking an acceleration mechanism more efficient than standard MHD acceleration.

\section{Possible additional observations}

To determine the radial dependences $B_*(r)$ and $N_*(r)$ along the jet, we need to measure the jet shape index $k$, which can be used to obtain $m=k\,b$ and $n=k\,s$. Alternatively, one can fit measurements of the core shift with the dependence $r\propto\nu^{-1/k_r}$ and use the fitted value of $k_r$ and Equation~(\ref{kr-k}) to obtain
\begin{equation}
k=k_r\,p\,.
\label{k_kr}
\end{equation} 
Thus, complementary measurements of core shifts (or a direct determination of a jet shape index $k$) enables one to infer how physical quantities $B_*$ and $N_*$ scale with distance $r$ along the jet.
In this case, the indices $m$ and $n$ can be expressed as functions of the observables $\{p,\,q,\,k_r\}$ and the assumed value of $t$:
\begin{equation}
m\,(p,\,q,\,k_\mathrm{r};\,t)=k_r(1+2q-pt),
\label{m}
\end{equation}
\begin{equation}
n\,(p,\,q,\,k_\mathrm{r};\,t)=k_r[(3-2\alpha)(pt-q)+p+1].
\label{n_index}
\end{equation}
Determination of $m$ and $n$ would allow extrapolation of the magnetic field strength and plasma number density to scales comparable to the gravitational radius. 

\begin{figure}
    \centering
    \includegraphics[width=1.0\linewidth]{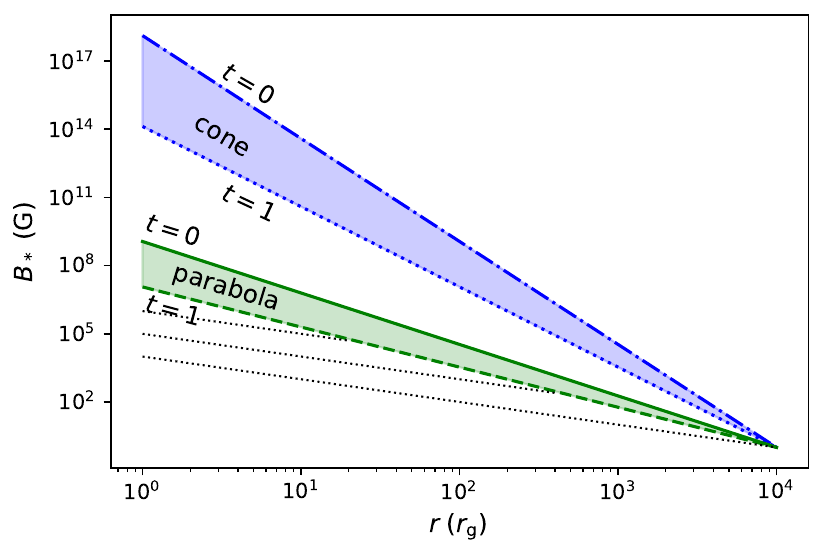}
    \caption{Illustration of potential impact of jet shape on estimates of magnetic field as a function of the de-projected distance along the jet, expressed in units of gravitational radius $r_\mathrm{g}$. We adopt the scaling 1~pc=$10^4\,r_\mathrm{g}$ (corresponding to a black hole mass of $M=5\times 10^9\,M_{\odot}$) and assume $B_* = 1$~G at a distance of 1~pc. Green lines and shaded region correspond to a parabolic jet with $k=0.5$, while blue lines and shaded region correspond to a conical jet with $k=1.0$. In each case, the shaded region covers different values of index $t$, ranging from $t=0$ (solid green and dash-dotted blue lines) to $t=1$ (dashed green and dotted blue lines). Black dotted lines show extrapolations assuming no magnetic flux dissipation up to $20\;r_\mathrm{g}$, $400\;r_\mathrm{g}$ and $10^4\;r_\mathrm{g}$.}
    \label{f:Bg}
\end{figure}

The impact of the jet shape index $k$ (or, reciprocally, the core shift index $k_r$) on the extrapolation of the jet rest-frame magnetic field down to gravitational-radius scales is illustrated in Fig.~\ref{f:Bg}, for a fiducial black hole with a mass  of $M=5\times10^9\,M_{\odot}$ and a magnetic field strength of 1~G at a distance of 1~pc. 

Adopting a conical jet shape, with $k=1$, implies a relatively large core shift index, $k_r\approx 1.6$, and results in extremely high values of $B_*(r_\mathrm{g})$. In contrast, a parabolic jet shape yields $k_r\approx 0.8$, consistent with the observed values \citep{Kutkin14,Ricci22,Perez_Diez25}, as well as magnetic field strength at the gravitational radius in the range of $10^{6.5}$--$10^{9.5}$~G. 

The lower end of this range can be reconciled, with some effort, with an accelerating parabolic flow generated in a system with a magnetized accretion disk around a supermassive black hole \citep[e.g.,][]{ZCST14}, where $B_*(r_\mathrm{g}) \approx 10^4\,\mathrm{G}\, (\eta_\mathrm{acc}/\eta_\mathrm{rad})^{1/2} (M_\mathrm{bh}/10^9\,\mathrm{M}_\odot)^{-1/2}$, and field strengths of $\sim 10^6\,\mathrm{G}$ can in principle be achieved, if the Eddington accretion efficiency, $\eta_\mathrm{acc}$, is substantially higher than the radiation efficiency, $\eta_\mathrm{rad}$. Even stronger horizon-scale fields are believed to be possible only for horizonless objects such as gravastars \citep{2001PhRvD..64j4022M} or wormholes \citep{2007IJMPD..16..909K}, and their detection could therefore be used for testing these exotic scenarios \citep{2017NatAs...1E..69L,2024evn..conf..101Z}. 

We note, however, that the inferred increase of the magnetic field toward the jet base may not persist all the way down to $r_\mathrm{g}$, but only to the scale at which magnetic flux dissipation ceases. If the magnetic flux becomes conserved below some radius, the magnetic field at $r_\mathrm{g}$ could be substantially lower than the values shown in Fig.~\ref{f:Bg}, depending on the location of the transition to flux conservation.

By analogy with quantum-mechanical terminology, one can propose a `complete set of observables' needed for making robust estimates of the basic properties of jets on ultracompact scales: measurements of the core width, brightness temperature, and position as functions of observing frequency. Such multifrequency data combined with the method presented here, enables the determination estimating the radial dependencies of the jet rest-frame emitting plasma number density $N_*(r)$, magnetic field $B_*(r)$ and width $d(r)$ (i.e. the jet boundary shape). The method does not assume any specific relation between $B_*$ and $N_*$. Moreover, it allows these physical quantities to be disentangled even in the absence of direct core shift measurements.

\section{Conclusions}

   \begin{enumerate}
      \item Multifrequency measurements of the core size and brightness temperature constrain the dependences of the jet rest-frame magnetic field $B_*$ and emitting plasma number density $N_*$ on the jet width $d$ under very general assumptions. The proposed method does not rely on any model-dependent relation between $B_*$ and $N_*$, thereby enabling a direct test of the equipartition assumption. 
      \item Additional measurements of either the core shift effect or the jet shape allow determination of the dependencies $B_*(r)$ and $N_*(r)$ along the jet. This, in turn, provides a means to extrapolate the magnetic field strength down to gravitational radius scales. 
      \item The only free parameter of the model is the exponent $t$, describing the dependence of the Doppler factor on jet width $\delta\propto d^{t}$. Physically, this parameter is expected to lie within the range $0\le t \le 1$.
      \item For the given values of $p$ and $q$ ($d\propto\nu^{-p}$, $T_\mathrm{br,\,obs}\propto\nu^{-q}$), the relation (Equation~\ref{Eq_s_b}) between the indices $b$ and $s$ ($B_*\propto d^{-b}$, $N_*\propto d^{-s}$) depends solely on the spectral index $\alpha$ determined by the energy distribution of emitting particles.
      \item The values of $p$ and $q$ reported by \citet{Roder25} pose tension with standard MHD jet models. The inferred value of $b$ from Equation~(\ref{b}) implies a magnetic field decline steeper than $B\propto d^{-2}$, expected for a purely poloidal magnetic field component under flux conservation. At the same time, the value of $s$ obtained from Equation~(\ref{s}) suggests the presence of continuous injection of emitting plasma (i.e. ongoing  `heating' and acceleration of cold plasma to ultra-relativistic energies) along the jet.
      \item The dependencies $B_*\propto d^{-3.5}$ and $N_*\propto d^{-0.6}$ indicate that the equipartition between the energy densities of the emitting plasma and the magnetic field is not maintained over the entire region of the jet probed in \citet{Roder25} by measurements at frequencies $\gtrsim 10\,\mathrm{GHz}$.
      \item The above results are based on quantities averaged for a heterogeneous sample of sources spanning a wide range of viewing angle (from BL Lac objects to radio galaxies). We therefore advocate conducting dedicated multifrequency measurements for individual sources to minimize potential biases associated with combining intrinsically diverse datasets.
   \end{enumerate}
   Based on these conclusions, we propose the use of a ``complete set of observables'' -- the core width, brightness temperature, and position measured across multiple frequencies -- for obtaining robust estimates of the radial dependencies $N_*(r)$, $B_*(r)$ and $d(r)$ via Equations~(\ref{k_kr}), (\ref{m}) and (\ref{n_index}). This approach does not require making any a priori assumption regardin the relation between $N_*$ and $B_*$ and relies only on observational constraints for the jet Doppler factor, which can be readily derived from VLBI data. Dedicated multifrequency measurements of three quantities will finally determination of plasma conditions in extragalactic relativistic jets.

\section*{Acknowledgements}

The authors thank Eduardo Ros for his valuable suggestions as well as the anonymous referee for the thoughtful comments that substantially improved this paper. E.E.N. and V.A.F. are supported in the framework of the State project ``Science'' by the Ministry of Science and Higher Education of the Russian Federation under the contract 075-15-2024-541. A.P.L. acknowledges support from M2FINDERS project which has received funding from the European Research Council (ERC) under the European Union’s Horizon 2020 research and innovation programme (grant agreement No 101018682).

\section*{Data availability}

There is no new data associated with the results presented in the paper. All the previously published data have the proper references.

\bibliographystyle{mnras}

\bibliography{nee1.bib} % your references Yourfile.bib

\bsp    % typesetting comment
\label{lastpage}

\end{document}